\documentclass[
preprint,
prl,
aps,
amsmath,
amssymb,
draft,
]{revtex4}

\def\half{\frac{1}{2}}

\def\slash#1{\, /\kern-0.6em{#1}}

\begin{document}

\title{Effect of a positive cosmological constant on cosmic
  strings}     
\author{Sourav Bhattacharya}\email{sbhatt@bose.res.in}
\author{Amitabha Lahiri}\email{amitabha@bose.res.in}
\affiliation{S. N. Bose National Centre for Basic Sciences, \\
Block JD, Sector III, Salt Lake, Calcutta 700 098, INDIA\\
}

\begin{abstract}
We study cosmic Nielsen-Olesen strings in space-times with a
positive cosmological constant. For the free cosmic
string in a cylindrically symmetric space-time, we calculate
the contribution of the cosmological constant to the angle
deficit, and to the bending of null geodesics. For a cosmic string
in a Schwarzschild-de Sitter space-time, we use Kruskal patches
around the inner and outer horizons to show that a thin string can
pierce them.
\end{abstract}

\maketitle

\section{1. Introduction}

In this paper, we study cosmic strings in space-times with a
positive cosmological constant. By cosmic string we will mean a
vortex line in the Abelian Higgs model~\cite{Nielsen:1973cs}. There
are several reasons for studying these. Recent observations suggest
a strong possibility that the universe is endowed with a positive
cosmological constant, $\Lambda>0\,$~\cite{Riess:1998cb,
Perlmutter:1998np}. This in turn implies that any observer will
find a cosmic horizon at a length scale of $\Lambda^{-\half}\,.$
Since $\Lambda$ is very small, and equivalently the length scale of
the horizon is very large, one might be tempted to neglect the
effect of $\Lambda$ on local physics. However, there are situations
in which local physics is affected by global topology.

An example comes from black hole no-hair statements. These say that
a static or stationary black hole is characterized by a small
number of parameters like its mass, angular momentum and charges
corresponding to long-range fields. These statements about the
uniqueness of black hole solutions were originally proven for
asymptotically flat space-times~\cite{Chrusciel:1994sn,
Heusler:1998ua}, but later extended to space-times with
$\Lambda>0$~\cite{Bhattacharya:2007zzb}. It was found that the
existence of a cosmic horizon introduced some subtleties in the
proofs of these statements. For an example which motivates us in
this paper, consider the Abelian Higgs model coupled minimally to
gravity. The only asymptotically flat, static, spherically
symmetric black hole has the Higgs field fixed at the minimum of
the potential, and the black hole is uncharged. On the other hand,
in the presence of a cosmic horizon, there is an additional black
hole solution in which the Higgs field is fixed at the maximum of
the potential and the black hole is charged~\cite{Bhattacharya:2007zzb}.
The black hole looks like the Reissner-N\"{o}rdstrom-de Sitter solution
with Higgs field in the false vacuum. This is of course the
opposite of the usual no-hair statement.

This new solution exists because of different boundary conditions
in the two cases --- in the asymptotically flat case these are
imposed at infinity, while for a positive cosmological constant it
is both convenient and sufficient to impose them at the cosmic
horizon. In general, given some asymptotically flat solution
(corresponding to $\Lambda=0$) of matter coupled to gravity, we may
find additional solutions, or qualitatively different ones, when
there is a cosmic horizon (corresponding to $\Lambda > 0$).

We are motivated by these arguments to look at infinitely long,
straight cosmic strings in space-times with $\Lambda>0\,.$ While
the role of such cosmic strings in cosmological perturbations and
structure formation is ruled out and the contribution of these
strings to the primordial perturbation spectrum must be less than
9\% (for a review and references
see~\cite{Perivolaropoulos:2005wa}), such strings could exist in
small numbers. How does a positive cosmological constant or a
cosmic horizon affect the physics of the string? It is known that a
string produces a conical geometry, or a deficit
angle~\cite{Vilenkin:1981zs}. It is also known that a cosmological constant
affects the bending of light~\cite{Rindler:2007zz,
Schucker:2007ut}. Both effects should be present in a string
space-time with $\Lambda> 0\,$.

We present analytical results for a cosmic string in two kinds of
space-time with a positive cosmological constant. The first is one
with only an infinite straight string, so that the space-time is
cylindrically symmetric with a string on the axis. We calculate the
angle deficit and the bending of light for this space-time. The
other space-time we consider is the Schwarzschild-de Sitter
space-time, and a cosmic string stretched between the inner and
outer horizons. For this we consider maximally extended (Kruskal)
coordinate patches near the horizons and find that the string can
extend beyond the horizons, i.e. pierce them.

\section{2. Free Cosmic String And Angle Deficit}
We start with the ansatz for a cylindrically symmetric static
metric~\cite{Vilenkin2:2000}
\begin{eqnarray}
ds^2=e^{A(\rho)}\left[-dt^2+ dz^2\right]+\rho^2 e^{B(\rho)}d\phi^2+ d\rho^2,
\label{metric}
\end{eqnarray}
and first solve for the cosmological constant vacuum
$R_{ab}-\frac{1}{2}R g_{ab}+ \Lambda g_{ab}=0$, with $\Lambda
>0$. There are three Killing vector fields here, the timelike
Killing field $(\partial_t)^a$ and two spacelike Killing fields $
(\partial_z)^a, (\partial_{\phi})^a$. The orbits of
$(\partial_{\phi})^a$ are closed spacelike curves which shrink to a
point as $\rho \rightarrow 0$.  We regard the set of points
$\rho=0$ as the axis of the space-time, and foliate this
space-time with $(\rho, \phi)$ planes (orthogonal to
$(\partial_t)^a$, $(\partial_z)^a$)\,.
A convenient coordinatization of these planes is by setting the
metric to be locally flat on the axis, i.e.
\begin{eqnarray}
ds^2\stackrel{\rho\rightarrow 0}\longrightarrow
-dt^2+dz^2+\rho^2d\phi^2+d\rho^2.  
\label{bcmetric}
\end{eqnarray}
We can always do this as long as there is no curvature singularity
on the axis. The vacuum solution subject to this boundary
condition is given by~\cite{Tian:1986, Linet:1986sr,
  BezerradeMello:2003ei} 
\begin{eqnarray}
ds^2=\cos^{\frac{4}{3}}\frac{\rho \sqrt{3\Lambda}}{2}
\left(-dt^2+dz^2 \right) +\frac{4}{3\Lambda}\sin^2 \frac{\rho
 \sqrt{3\Lambda}}{2}\cos^{-\frac{2}{3}}\frac{\rho
 \sqrt{3\Lambda}}{2}d\phi^2+d\rho^2.
\label{vacuum}
\end{eqnarray}
The metric is singular at $\rho=\frac{n\pi}{\sqrt{3\Lambda}}$,
where $n$ are integers. Of these points, those corresponding to
even $n$ are flat, with $n=0$ being the axis. The points
corresponding to odd $n$ are curvature singularities. The quadratic
invariant of Riemann tensor behaves there as
\begin{eqnarray}
R_{abcd}R^{abcd} \approx \frac{\Lambda^2}{\left(\frac{n\pi}{2}-
 \frac{\rho \sqrt{3\Lambda}}{2}\right)^4}\,,  n~\text{odd}.  
\label{curvature}
\end{eqnarray}
The curvature singularity at $n=1$ is not protected by a horizon, 
so it must be a naked singularity. The singularities for higher $n$
thus appears to be unphysical, and will not concern us further.

Our region of interest will be near the axis and far from the naked
singularity at $n=1$. In this region, we consider Einstein's
equations with energy-momentum tensor, $R_{ab}-\frac{1}{2}R
g_{ab}+\Lambda g_{ab}=8 \pi G T_{ab} $, The energy-momentum tensor
$T_{ab}$ corresponds to that of a string solution of the Abelian
Higgs model, which has the Lagrangian
\begin{eqnarray}
{\cal{L}} = -\left(D_{a}\Phi\right)^{\dagger}\left(D^{a}\Phi\right)
- \frac{1}{4}\tilde{F}_{ab}\tilde{F}^{ab} -
\frac{\lambda}{4}\left(\Phi^{\dagger}\Phi - \eta^2\right)^2.
\label{lagrangian}
\end{eqnarray}
Here~$D_{a}=\nabla_{a}+ieA_{a}$~is the gauge covariant
derivative,~$\tilde{F}_{ab}=\nabla_{a}A_{b}-\nabla_{b}A_{a}$ is the
electromagnetic field strength tensor and $\Phi$ is a complex
scalar. For convenience of calculations we will parametrize $\Phi$
and $A_{a}$ as
\begin{eqnarray}
\Phi=\eta X e^{i\chi}, \qquad
A_{a}=\frac{1}{e}\left[P_{a}-\nabla_{a}\chi\right].
\label{gaugecov}
\end{eqnarray}
The Abelian Higgs model has string like solutions in flat
space-time~\cite{Nielsen:1973cs}.  For these solutions, the phase
$\chi$ is multiple valued outside the string,
\begin{eqnarray}
\oint \nabla_a \chi d x^a=\oint d \chi=2n\pi,
\label{multiphase}
\end{eqnarray}
where the integral is done over a closed loop around the string and
$n\,$ called the winding number, is some nonzero integer called the
winding number. On the other hand, $\chi$ is single valued inside
the core, so the Lagrangian inside the core becomes
\begin{eqnarray}
{\cal{L}}=-\eta^2 \nabla_{a}X\nabla^a X -\eta^2 X^2 P_a P^a
-\frac{1}{4 e^2} F_{ab}F^{ab}-\frac{\lambda \eta^4}{4}
\left(X^2-1\right)^2,
\label{lagrangian2}
\end{eqnarray}
where $F_{ab}=\nabla_{a} P_b -\nabla_{b}P_a$. We will write
$\rho_0$ for the core radius. Due to the cylindrical symmetry of
the space-time we can take the following ansatze for $X$ and $P_a$
\begin{eqnarray}
X=X(\rho)\,;\qquad P_a=P(\rho)\nabla_a \phi\,.
\label{fieldansatz}
\end{eqnarray}
The energy-momentum tensor is taken to be non-zero only inside the
string core, and zero outside .  The various non-vanishing
components of energy momentum tensor $T_{ab}$ for the Abelian Higgs
model in cylindrical coordinates (\ref{metric}) are
\begin{eqnarray}
&T_{tt}&=\left[\eta^2 {X^{\prime}}^2 + \frac {\eta^2 X^2 P^2
     e^{-B}}{\rho^2}+\frac{{P^{\prime}}^2 e^{-B}}{2 e^2 \rho^2}
   +\frac{\lambda \eta^4}{4} \left(X^2-1\right)^2 \right]
 e^A. \nonumber \\  
&T_{\rho\rho}&= \left[\eta^2 {X^{\prime}}^2-\frac {\eta^2 X^2 P^2
     e^{-B}}{\rho^2}+\frac{{P^{\prime}}^2 e^{-B}}{2 e^2 \rho^2}
     -\frac{\lambda \eta^4}{4} \left(X^2-1\right)^2
     \right]. \nonumber \\  
&T_{\phi \phi}&=\left[-\eta^2 {X^{\prime}}^2 + \frac {\eta^2 X^2
     P^2 e^{-B}}{\rho^2}+\frac{{P^{\prime}}^2 e^{-B}}{2 e^2
     \rho^2} -\frac{\lambda \eta^4}{4} \left(X^2-1\right)^2
     \right] \rho^2 e^B. \nonumber \\  
&T_{zz}&= -\left[\eta^2 {X^{\prime}}^2 + \frac {\eta^2 X^2 P^2
     e^{-B}}{\rho^2}+\frac{ {P^{\prime}}^2 e^{-B}}{2 e^2 \rho^2}
     +\frac{\lambda \eta^4}{4} \left(X^2-1\right)^2 \right]
     e^A. 
\label{emtensor}
\end{eqnarray}
Since $\Phi=\eta X e^{i \chi}$, we have along a closed loop of
$X=\text{constant}$ outside the string core  
\begin{eqnarray}
\oint d \chi=2n\pi= \frac{1}{i}\oint \frac{d \Phi}{\Phi}\,.
\label{phase}
\end{eqnarray}
It is clear that $\Phi=0$ somewhere inside the loop and hence $X=0$
somewhere inside the loop. For the string solution the Higgs field
should vanish as we approach the axis, and should approach its
vacuum expectation value outside the string. The gauge field
$A_{\phi}$ should accordingly approach
$-\frac{1}{e}\partial_{\phi}\chi$ away from the string and a
constant on the axis. In other words, $X\rightarrow 0$, $P
\rightarrow 1$ as we approach the axis, while $X\rightarrow 1$, $P
\rightarrow 0$ outside the string core.

We now return to Einstein equations $R_{ab}-\frac{1}{2}R
g_{ab}+\Lambda g_{ab}=8\pi G T_{ab}$. The variation of $X$ and
$P\,,$ and thus of the energy-momentum tensor, near the `string
surface' at $\rho = \rho_0$ is a problem of considerable interest
and has been studied numerically in various papers. However, in
this paper we are concerned about the existence of the string and
its behavior near the horizons. Accordingly, we will fix boundary
conditions by assuming $X=0$, $P=1$ inside the string core and
$X=1$, $P=0$ outside. This guarantees that the energy-momentum
tensor (\ref{emtensor}) is identically zero outside the core. The
fields are assumed to be smoothed out at the string
surface at $\rho=\rho_0$ such that the local conservation
law $\nabla_{a}T^{ab}=0$ remains valid. Then we can solve Einstein
equations to find inside the core $(0\leq \rho <\rho_0)$
\begin{eqnarray}
ds^2\approx\cos^{\frac{4}{3}}\frac{\rho
\sqrt{3\Lambda^{\prime}}}{2}\left(-dt^2+dz^2\right)
+\frac{4}{3\Lambda^{\prime}}\sin^2 \frac{\rho
\sqrt{3\Lambda^{\prime}}}{2}\cos^{-\frac{2}{3}}\frac{\rho
\sqrt{3\Lambda^{\prime}}}{2}d\phi^2+d\rho^2.    
\label{metric3}
\end{eqnarray}
This solution inside the string is the same as the vacuum solution
of Eq.~(\ref{vacuum}), but with a modified cosmological constant
\begin{eqnarray}
\Lambda^{\prime}=\Lambda+2\pi G \lambda \eta^{4}\,. 
\label{lambdaprime}
\end{eqnarray}
 The solution
for the metric in the vacuum region outside the string is given by
\begin{eqnarray}
ds^2=\cos^{\frac{4}{3}}\frac{\rho
 \sqrt{3\Lambda}}{2}\left(-dt^2+dz^2\right)
+\delta^2\frac{4}{3\Lambda}\sin^2 \frac{\rho
 \sqrt{3\Lambda}}{2}\cos^{-\frac{2}{3}}\frac{\rho
 \sqrt{3\Lambda}}{2} d\phi^2+d\rho^2.  
\label{metric2}
\end{eqnarray}
This solution differs from the vacuum solution by the presence of a
number $\delta$, which is related to the angle deficit.
In~\cite{Ghezelbash:2002cc}, where vortices in de Sitter space was
studied perturbatively, the authors argued for the existence of
this $\delta$, but did not estimate it. Here we evaluate $\delta$
in the following way.  We first compute
\begin{eqnarray}
\frac{1}{2 \pi}
\int \int\sqrt{g^{(2)}} d\rho d\phi 
\left(G_{t}\,^{t}+\Lambda\right)
\label{gbonet}
\end{eqnarray}
on $(\rho,\phi)$ planes. Here $g^{(2)}$ is the determinant of the
induced metric on these planes. It is clear that $\delta$ appears
due to the energy-momentum tensor which is confined to the region
$\rho \leq \rho_0$.  Then calculating $G_{t}\,^{t}$ from
the general ansatz (\ref{metric}), we have
\begin{eqnarray}
\int_{0}^{\rho_0}\sqrt{g^{(2)}} d\rho 
\left(G_{t}\,^{t}+\Lambda\right)=
\int_{0}^{\rho_0}d\rho\left[
\rho e^{\frac{B}{2}}\left(\frac{{A^{\prime}}^2}{4} + \Lambda\right) 
+ \left(\rho e^{\frac{B}{2}}\frac{{A^{\prime}}}{2}\right)' +
\left(\rho e^{\frac{B}{2}}\right)'' \right],
\label{mucalc}
\end{eqnarray}
where a prime denotes differentiation with respect to $\rho$.
But according to Einstein equation, $G_{t}\,^{t}+\Lambda=8
 \pi G T_{t}\,^{t}$. Substituting the value of 
$G_{t}\,^{t}+\Lambda$ in Eq. (\ref{mucalc}),
we get
\begin{eqnarray}
 \frac{d}{d\rho}\left(\rho e^{\frac{B}{2}}\right)\Bigg\vert^{\rho_0}_{0}+
 \left(\rho e^{\frac{B}{2}}\frac{A'}{2}
 \right)\Bigg\vert^{\rho_0}_{0}=-4G\mu - \int_{0}^{\rho_{0}}d\rho
 \rho e^{\frac{B}{2}}\left(\Lambda +\frac{{A^{\prime}}^2}{4}\right), 
\label{mu2}
\end{eqnarray}
where
\begin{eqnarray}
  \mu:=-\int_0^{2\pi}\int_0^{\rho_0}  d\phi  d\rho
\rho e^{\frac{B}{2}}T_{t}\,^{t}\approx\frac{\pi \lambda
  \eta^4}{\Lambda^{\prime}}\left[1-\cos^{\frac{2}{3}} \frac{\rho_0
    \sqrt{3 \Lambda^{\prime}}}{2}\right] 
\label{mu} 
\end{eqnarray}
is the string mass per unit length. To get the approximate expression
for $\mu$ in Eq. (\ref{mu}) we have used $T_{t}\,^{t}=
-\frac{\lambda \eta^4}{4}$ which is due to our approximation
$X=0$ and $P=1$ inside the core. Outside the core $T_{t}\,^{t}=0$
identically, so we have used the metric functions in 
Eq. (\ref{metric3}). In evaluating the total
derivative terms in the left hand side of Eq.~(\ref{mu2}), we will
use the interior metric of Eq.~(\ref{metric3}) at $\rho=0$, but the
vacuum metric of Eq.~(\ref{metric2}) at the string `surface'
$\rho=\rho_0$. This
requires an explanation.  If we assume the energy-momentum to be
non-vanishing only within the string, the right hand side of
Eq.~(\ref{mucalc}) will have contributions only from $\rho\leq
\rho_0\,,$ as we have written. The integrand on the left hand side
Eq.~(\ref{mucalc}) also vanishes outside the string according to vacuum
Einstein equations.  When we integrate the left hand side, we
should do so to the surface of the string, i.e., where the
energy-momentum tensor vanishes.
 But at that point we have the
vacuum solution of Eq.~(\ref{metric2}), so that is what we should
use at the upper limit of integration.  Thus we find
\begin{eqnarray}
 1-\delta \left(\cos^{\frac{2}{3}}\frac{\rho_0 \sqrt{3
 \Lambda}}{2}-\frac{1}{3}\cos^{-\frac{4}{3}}\frac{\rho_0 \sqrt{3
 \Lambda}}{2} \sin^2 \frac{\rho_0 \sqrt{3 \Lambda}}{2}\right)=
 4G\mu + \int_{0}^{\rho_{0}}d\rho \rho e^{\frac{B}{2}}\left(\Lambda
 +\frac{{A^{\prime}}^2}{4}\right)\,.\, 
\label{delta}
\end{eqnarray}

The integrals on the right hand side of Eq.~(\ref{delta}) cannot be
evaluated explicitly, since the integrand cannot be written as a
total derivative, nor do we know the detailed behavior of the
metric near the string surface at $\rho=\rho_0\,.$ However, we can
make an estimate of these integrals using the value of the metric
coefficients inside the core. This means that we ignore the details
of the fall off of the energy-momentum tensor near $\rho =
\rho_0\,.$ Then using Eq.~(\ref{metric3}) we get from
Eq.~(\ref{delta}) an expression for $\delta$
\begin{eqnarray}
\delta= \frac{1-4G\mu-\frac{2\Lambda}{\Lambda^{\prime}}
 \left(1-\cos^{\frac{2}{3}}\frac{\rho_0 
\sqrt{3\Lambda^{\prime}}}{2} \right)
+\frac{1}{3}\left(1-\cos^{-\frac{4}{3}}\frac{\rho_0
 \sqrt{3\Lambda^{\prime}}}{2} \right)
+\frac{2}{3}\left(1-\cos^{\frac{2}{3}}\frac{\rho_0
 \sqrt{3\Lambda^{\prime}}}{2} \right)}
{\left(\cos^{\frac{2}{3}}\frac{\rho_0 \sqrt{3
\Lambda}}{2}-\frac{1}{3}\cos^{-\frac{4}{3}}\frac{\rho_0 \sqrt{3
\Lambda}}{2} \sin^2 \frac{\rho_0 \sqrt{3 \Lambda}}{2}\right)}\,.
\label{deltafinal}
\end{eqnarray}
This result may be compared with one obtained in~\cite{Deser:1983dr}
where the authors considered point particles as source and solved
Einstein equations in $2+1$ dimensional de Sitter space. The `particles'
may be thought of as punctures created in the plane by infinitely thin 
long strings. 
A conical singularity was found, with an angle deficit
$\delta=(1-4 G m)$ where $m$ is the mass of
the particle. Our result includes corrections dependent on $\Lambda$, 
which we may think of as coming from the finite thickness of the string.
The size of the string $\rho_0$ is of the order of $(\sqrt\lambda
\eta)^{-1}\,,$ at least when the winding number is
small~\cite{Brihaye:2008uy}. This is essentially because the metric
is flat on the axis, so we can approximate $\rho_0$ by its value in
flat space. Further, the scale of symmetry breaking $\eta$ is small
compared to the Planck scale in theories of particle physics in
which cosmic strings appear. For example, the grand unified scale
is about $10^{16}$ GeV, so that $G\eta^2 \sim 10^{-3}\,.$ It is
also reasonable to assume that the string size is small compared to
the cosmic horizon. In other words, we assume $\rho_0^2\Lambda \ll
1\,.$ Thus we find that $\rho_0^2\Lambda' \ll 1\,$ as well, where
$\Lambda'$ is given by Eq.~(\ref{lambdaprime})\,.

Then by expanding $G\mu$ using the expression in Eq.~(\ref{mu}), we
find $\mu = \frac\pi4 \lambda\eta^4\rho_0^2$ approximately, and thus
$G\mu \ll 1\,.$ We can also find an approximate expression for
$\delta$ from Eq.~(\ref{deltafinal}) under these assumptions, 
\begin{eqnarray}
\delta \approx 1 - 4G\mu\left(1 + \frac34\rho_0^2\Lambda +
G\mu\right)\,.   
\label{deltacorr}
\end{eqnarray}
The leading correction to $\delta$ due to the cosmological constant
is of a higher order of smallness, as we can see from this. The
meaning of $\delta$ is obvious in space-times with vanishing
cosmological constant, for which Eq.~(\ref{delta}) was worked out
in~\cite{Garfinkle:1985hr} to find $\delta\approx 1 - 4G\mu\,,$
where $G\mu\ll 1$ as before, and ${\cal O}(G^2\mu^2)$ corrections
were ignored. Then asymptotically we get the conical space-time
\begin{eqnarray}
ds^2=-dt^2+d\rho^2+dz^2+\rho^2\delta^2 d\phi^2.
\label{levicivita}
\end{eqnarray}
In this space-time the azimuthal angle runs from $0$ to
$2\pi\delta\,.$ So Eq.~(\ref{levicivita}) is Minkowski space-time
minus a wedge which corresponds to a deficit $2\pi (1-\delta)\,$ in
the azimuthal angle. The difference of initial and final azimuthal
angles of a null geodesic (i.e., light ray in the geometrical
optics approximation) at $\rho \rightarrow \infty$ is
$\frac{\pi}{\delta}$~\cite{Vilenkin:1981zs, Vilenkin2:2000}.  Light
bends towards the string even though the curvature of space-time is
zero away from the axis.

For a positive cosmological constant, the metric in the exterior of
the string is 
\begin{eqnarray}
ds^2=\cos^{\frac{4}{3}}\frac{\rho
 \sqrt{3\Lambda}}{2}\left(-dt^2+dz^2\right) +
\delta^2 \frac{4}{3\Lambda}\sin^2 \frac{\rho
 \sqrt{3\Lambda}}{2}\cos^{-\frac{2}{3}}\frac{\rho
 \sqrt{3\Lambda}}{2} d\phi^2+d\rho^2\,,
\label{metricfinal}
\end{eqnarray}
with $\delta$ given in Eq.~(\ref{deltafinal}) or
Eq.~(\ref{deltacorr}). Comparing with the string-free vacuum
solution of Eq.~(\ref{vacuum}) we find that, similar to the
asymptotically flat space-time, the deficit in the azimuthal angle
in space-time with a positive cosmological constant is also
$2\pi(1-\delta)\,,$ but now with $\delta$ given by
Eq.~(\ref{deltacorr}). However the bending of null geodesics will
be quite different in the cosmic string space-time of
(\ref{metricfinal}) from that in asymptotically flat cosmic string
space-time. 

Since our space-time (\ref{metricfinal}) has a translational
isometry along $z\,,$ for the sake of simplicity we can consider
null geodesics on the $z=0$ plane. It is well known that if
$\chi^a$ is a Killing field, then for any
geodesic with tangent $u^a$,
the quantity $g_{ab}u^a \chi^b$ is conserved along the geodesic.
Thus the conserved angular momentum of a future directed 
null test particle in the space-time (\ref{metricfinal}) is
\begin{eqnarray} 
L= g_{ab}(\partial_\phi)^a u^b
=\delta^2
\frac{4}{3\Lambda}\sin^2 \frac{\rho
\sqrt{3\Lambda}}{2}\cos^{-\frac{2}{3}}\frac{\rho
\sqrt{3\Lambda}}{2} \dot{\phi},
\label{angularmom}
\end{eqnarray}
 while its conserved energy is
\begin{eqnarray}
 E=-g_{ab}(\partial_t)^a u^b
=\cos^{\frac{4}{3}}\frac{\rho \sqrt{3\Lambda}}{2} \dot{t}.
\label{energy}
\end{eqnarray}
The dot denotes differentiation with respect to an
affine parameter and $(\partial_\phi)^a$ and
$(\partial_t)^a$ are rotational and time translational Killing
fields respectively.
Then for null geodesics on the $z=0$ plane it is straightforward to
obtain
\begin{eqnarray}
{\frac{d\phi}{d\rho}}={\frac{3\Lambda L}{4  E \delta^2}
 \frac{\cos^{\frac{4}{3}}\frac{\rho\sqrt{3\Lambda}}{2}}
      {\sin^2\frac{\rho \sqrt{3\Lambda}}{2} \left[1-\frac{3
         \Lambda L^2}{4 E^2 \delta ^2 } \cot^2 \frac{\rho
         \sqrt{3\Lambda}}{2}\right]^{\frac{1}{2}}} }~.  
\label{bending1}
\end{eqnarray}
Since $(\rho, \phi)$ are smooth functions of the affine parameter
the derivative on the left hand side of Eq. (\ref{bending1}) is well
defined. From the null geodesic equation on $z=0$ plane we have the
distance of closest approach to the string
\begin{eqnarray}
\rho_c=\frac{\, 2}{\sqrt{3 \Lambda}}\tan^{-1}
\frac{\sqrt{3\Lambda }L}{2E \delta}~.
\label{impact}
\end{eqnarray}
Let us consider a null geodesic which starts from some point
$(\rho_{\text{max}}, \phi_1)$ between the string surface and the 
singularity at $\rho= \frac{\pi}{\sqrt{3 \Lambda}}$. We will
look at till it reaches a point $(\rho_{\text{max}}, \phi_2)$.
For simplicity of interpretation, we have chosen the `initial'
and `final' radial distances to be equal. Since the trajectory of
the geodesic will be symmetric about the distance of closest 
approach $\rho_c$, the change in the azimuthal angle is
\begin{eqnarray}
\Delta \phi=\phi_2-\phi_1
={\frac{3\Lambda L}{2  E \delta^2}
\int_{\rho_c}^{\rho_{\text{max}}}
 \frac{\cos^{\frac{4}{3}}\frac{\rho\sqrt{3\Lambda}}{2}}
      {\sin^2\frac{\rho \sqrt{3\Lambda}}{2} \left[1-\frac{3
         \Lambda L^2}{4 E^2 \delta ^2 } \cot^2 \frac{\rho
         \sqrt{3\Lambda}}{2}\right]^{\frac{1}{2}}} }d\rho~.  
\label{bending2}
\end{eqnarray}
Eq. (\ref{bending2}) along with the expression for $\rho_c$
determines the change of $\phi$ with $\rho$.  The full expression
for the integral in Eq. (\ref{bending2}) is rather messy and we
will look at two special cases only.  First, when $\rho$ is much
smaller than the radius of the cosmological singularity $\left(\rho
\ll \frac{\pi}{\sqrt{3\Lambda}} \right)$, we have approximately
\begin{eqnarray}
\Delta \phi\approx \frac{2}{\delta}
\sec^{-1}\left(\sqrt{1 + k^2}    
\frac{\rho E\delta}{L} \right)
\Bigg\vert_{\rho_c}^{\rho_{\text{max}}}
 - \frac{4 k}  {3 \delta\sqrt{1 + k^2}} 
\left(\frac{\rho^2 3 \Lambda}{4}-\frac{k^2}{1 + k^2}
\right)^{\frac{1}{2} }
\Bigg\vert_{\rho_c}^{\rho_{\text{max}}},
\label{bending3}
\end{eqnarray}
where $k = \frac {\sqrt{3\Lambda} L} {2 E \delta}$.  The second
term in Eq. (\ref{bending3}) is negative and the repulsive effect
of positive $\Lambda$ is manifest in this term. In the $\Lambda
\rightarrow 0$ limit only the first term survives.  In that case
the limit $\rho_{\text{max}} \rightarrow \infty$ recovers the well
known formula $\Delta \phi= \frac{\pi}{\delta}$.  Next, near the
singularity $\rho \rightarrow \frac{\pi}{\sqrt{3 \Lambda}}$, we can
approximately write $\cos \frac{\rho\sqrt{3\Lambda} } {2}\approx
\left( \frac{\pi}{2}-\frac{\rho\sqrt{3\Lambda}}{2} \right)$ and
integrate Eq. (\ref{bending2}) to get
\begin{eqnarray}
\Delta \phi\approx -\frac {6 k}{\delta}
\left[\frac{1}{7}\left(\frac{\pi}{2}- \frac{\rho
 \sqrt{3
     \Lambda}}{2}\right)^{\frac{7}{3}
 }+\frac{k^2}{26}\left(\frac{\pi}{2}- \frac{\rho
 \sqrt{3
     \Lambda}}{2}\right)^{\frac{13}{3} }+\dots\right]
_{\rho_c}^{\rho_{\text{max}}}.
\label{bending4}
\end{eqnarray}
%

\section{3. Black Hole Pierced By A String}
If a cosmic string pierces a Schwarzschild black hole, the
resulting space-time has a conical singularityn~\cite{Aryal:1986sz}
as well.  The authors of~\cite{Achucarro:1995nu} showed by
considering the equations of motion of the matter fields that an
Abelian Higgs string (for both self gravitating and non
self-gravitating matter) can pierce a Schwarzschild black hole. We
will adapt in this section the method described
in~\cite{Achucarro:1995nu} to establish that a Schwarzschild-de
Sitter black hole can be similarly pierced by a Nielsen-Olesen
string.

Inside the core of a string, we can derive the equations of motion
for  the fields from (\ref{lagrangian2}),
\begin{eqnarray}
\nabla_{a}\nabla^{a}X- X P_{a}P^{a}-\frac{\lambda
 \eta^2}{2}X\left(X^2-1\right)=0,
\label{eom1}
\end{eqnarray}
\begin{eqnarray}
\nabla_{a}F^{ab}-2 e^2 \eta^2 X^2 P^{b}=0.
\label{eom2}
\end{eqnarray}
Consider for a moment flat space cylindrical coordinates $(t, \rho,
\phi, z)$ and take the scalar field $X$ to be cylindrically
symmetric, $X=X(\rho)$. Also assume that the gauge field $P_a$ can
be written as $P_{a}=P(\rho)\nabla_{a}\phi$.  Then the equations of
motion (\ref{eom1}) and (\ref{eom2}) become
\begin{eqnarray}
\frac{d^2X}{d \rho^2}+\frac{1}{\rho}\frac{dX}{d\rho}-                                              
\frac{X P^2}{\rho^2} -\frac{X}{2}(X^2-1)=0,
\label{noeq1}
\end{eqnarray}
\begin{eqnarray}
\frac{d^2 P}{d \rho^2}-\frac{1}{\rho}\frac{dP}{d\rho}-                                             
\frac{2 e^2}{\lambda} X^2 P=0.
\label{noeq2}
\end{eqnarray}
Here we have scaled $\rho$ by $\left(\sqrt{\lambda}\eta
\right)^{-1}$ to convert it to a dimensionless radial
coordinate. These are the usual equations which were shown
in~\cite{Nielsen:1973cs} to have string like solutions.  We wish to
show that these equations hold also in the Schwarzschild-de Sitter
background space-time if the string thickness is small compared to
the black hole event horizon, and if we neglect the backreaction of
the string on the metric.

The Schwarzschid-de Sitter metric in the usual spherical polar
coordinates reads
\begin{eqnarray}
ds^2= -\left(1-\frac{2M}{r}-\frac{\Lambda r^2}{3}\right)dt^2+
\left(1-\frac{2M}{r}-\frac{\Lambda
r^2}{3}\right)^{-1}dr^2+r^2d\Omega^2.
\label{sdsmetric}
\end{eqnarray}
There are three horizons in this space-time $-$ the black hole
event horizon at $r=r_{H}$, the cosmological horizon at $r=r_{C}$
and an unphysical horizon at $r=r_U$ with $r_U<0$. If we assume
$2M\ll \frac{1}{\sqrt{\Lambda}}$, which we will throughout, we find
that the approximate sizes of the horizons are $r_H\approx 2M$,
$r_C \approx \sqrt{\frac{3}{\Lambda}}$ and $r_U \approx
-\sqrt{\frac{3}{\Lambda}}$. The string we are looking for is thin
compared to the horizon size, i.e. we will assume also that
\begin{eqnarray}
\frac{1}{\sqrt{\lambda}\eta} \ll 2M \ll \frac{1}{\sqrt \Lambda}~.
\label{approx}
\end{eqnarray}
We now expand the field equations in the Schwarzschild-de Sitter
background; in other words, we will neglect the backreaction on the
metric due to the string. Then Eq.~(\ref{eom1}) becomes
\begin{eqnarray}
\frac{1}{r^2} \partial_{r} \left[r^2
 \left(1-\frac{2M}{r}-\frac{\Lambda r^2}{3}\right)
 \partial_{r}X\right] +\frac{1}{r^2 \sin^2
 \theta}\partial_{\theta}\left(\sin \theta \partial_{\theta}X
\right) -\frac{X P^2}{r^2 \sin^2 \theta} -\frac{\lambda \eta^2}{2}
X\left(X-1\right)=0.\nonumber \\
\label{scalareq}
\end{eqnarray}
For the string solution the matter distribution will be
cylindrically symmetric. For convenience we take the string
along the axis $\theta=0$, although our arguments are clearly valid for
$\theta=\pi$ as well. We define
as before a dimensionless cylindrical radial coordinate $\rho =
r\sqrt{\lambda}{\eta}\sin\theta$. For cylindrically symmetric
matter distribution both $(X, P)$ will be functions of $\rho$
only. With this we can rewrite (\ref{scalareq}) as
\begin{eqnarray}
\left(\sin^2\theta -\frac{2M \sqrt{\lambda}\eta
 \sin^3\theta}{\rho}-\frac{\overline{\Lambda} \rho^2}{3}\right)
\left[\frac{d^2 X}{d\rho^2}+\frac{2}{\rho}\frac{d X}{d\rho}\right]+
\left(\frac{2M \sqrt{\lambda}\eta \sin^3\theta}{\rho^2}-\frac{2
 \overline{\Lambda} \rho}{3}\right) \frac{dX}{d\rho}&+&\nonumber
\\  
\left[ \frac{1}{\rho}\frac{dX}{d\rho}\cos^2 \theta
 -\frac{1}{\rho}\frac{dX}{d\rho} \sin^2\theta+
 \frac{d^2X}{d\rho^2} \cos^2\theta \right]-\frac{X
 P^2}{\rho^2}-\frac{1}{2} X\left(X-1\right)=0&,&
\label{scalareq2}
\end{eqnarray}
where $\overline {\Lambda}=\frac{\Lambda}{\lambda \eta^2}$ is a
dimensionless number. Under the approximations of
Eq.~(\ref{approx}) and because $\sin \theta \ll 1 $ inside the
string core, Eq.~(\ref{scalareq2}) reduces to Eq.~(\ref{noeq1}),
i.e., the flat space equation of motion for Abelian Higgs
model. Outside the string core $(\rho>1)$, we can as before set
$X=1$ . Hence we can say that Eq.~(\ref{scalareq2}), and hence
Eq.~(\ref{scalareq}) gives rise to a configuration of scalar field
$X$ similar to that of the Nielsen-Olesen string. A similar
calculation for Eq.~(\ref {eom2}) in the Schwarzschild- de Sitter
background shows that it reduces to Eq.~(\ref{noeq2}) under the
same approximations. Thus we conclude that the Schwarzschild-de
Sitter space-time allows a uniform Nielsen-Olesen string along
the axis $\theta=0$ in the region $r_{H}< r < r_{C}$. 

From the calculations so far we cannot conclude how the string behaves
at or near the horizons. The reason is the following.
The two horizons at $(r_{H}, r_{C})$
appear as two coordinate singularities in the chart
described in Eq. (\ref{sdsmetric}). Clearly we cannot expand
the field equations in this coordinate system at or around the 
horizons. To do the expansion we need to use maximally
extended coordinates which will be free from coordinate 
singularities for the Schwarzschild-de Sitter space-time, and 
has only the curvature singularity at $r=0$. Let us construct
Kruskal like patches at the two horizons to remove 
the two coordinate singularities.

First we consider radial null geodesics in the vicinity of the black hole
event horizon $r_H$. For these geodesics  
\begin{eqnarray}
\frac{dt}{dr}=\pm \frac{1}{\left(1-\frac{2M}{r}-\frac{\Lambda
   r^2}{3}\right)}=\pm \frac{3r}{\Lambda
 \left(r-r_H\right)\left(r-r_U\right)\left(r-r_C\right)}~.
\label{extension1}
\end{eqnarray}
Eq.~(\ref{extension1}) can be easily integrated to give
\begin{eqnarray}
t=\pm r_{*}+\text{constant},
\label{extension2}
\end{eqnarray}
where $r_{*}$ is the tortoise coordinate given by
\begin{eqnarray}
r_{*}=\alpha \ln \left \vert \frac{r}{r_H}-1\right\vert + \beta
\ln \left \vert\frac{r}{r_C}-1\right\vert + \gamma \ln \left \vert
\frac{r}{r_U}-1\right\vert~.  
\label{extension3}
\end{eqnarray}
The three constants  $(\alpha, \beta, \gamma) $ are given by
\begin{eqnarray}
\alpha=\frac{3
r_{H}}{\Lambda\left(r_{C}-r_{H}\right)\left(r_{H}-r_U\right)},\quad
\beta=-\frac{3
r_{C}}{\Lambda\left(r_{C}-r_H\right)\left(r_{C}-r_U\right)},\quad
\gamma=-\frac{3r_U}{\Lambda\left(r_{C}-r_U\right)\left(r_H-r_{U}\right)}~.
\nonumber\\  
\label{kruskal2}
\end{eqnarray}
In ($t$, $r_{*}$) coordinates the radial metric becomes
\begin{eqnarray}
ds^2_{\text{radial}}=\left(1-\frac{2M}{r}-\frac{\Lambda
 r^2}{3}\right)\left(-dt^2+dr_{*}^{2} \right).
\label{extension4}
\end{eqnarray}

Defining null coordinates ($u$, $v$) such that
\begin{eqnarray}
u=t-r_{*} \qquad\text{and}\qquad v=t+r_{*},
\label{extension5}
\end{eqnarray}
we can write the radial metric (\ref{extension4}) as
\begin{eqnarray}
ds^2_{\text{radial}}=-\left(1-\frac{2M}{r}-\frac{\Lambda r^2}{3}\right)du dv.
\label{extension6}
\end{eqnarray}
Now we define timelike and spacelike coordinates ($T$, $Y$) by
\begin{eqnarray}
T:=\frac{e^{\frac{v}{2\alpha}} -e^{-\frac{u}{2\alpha}} }{2};\qquad
Y:= \frac{e^{\frac{v}{2\alpha}} +e^{-\frac{u}{2\alpha}} }{2}~.
\label{extension7}
\end{eqnarray}
($T$, $Y$) satisfy the following relations
\begin{eqnarray}
Y^2-T^2&=&\left[\frac{r}{r_H}-1 \right]
e^{\frac{\beta}{\alpha}\ln\left\vert\frac{r}{r_{C}} - 1\right\vert +
 \frac{\gamma}{\alpha}\ln\left\vert\frac{r}{r_U}-1\right\vert},  
\label{kruskal3}
\end {eqnarray}
\begin{eqnarray}
\frac{T}{Y}&=&\tanh \left(\frac{t}{2\alpha}\right).
\label{extension8}
\end {eqnarray}
In terms of ($T$, $Y$), the full space-time metric of
Eq.~(\ref{sdsmetric}) becomes
\begin{eqnarray}
ds^2=\frac{8M\alpha^2}{r} \left\vert\frac{r}{r_U}-1 \right\vert^
{1-\frac{\gamma}{\alpha}}\left\vert\frac{r}{r_{C}}-1 \right\vert^
{1-\frac{\beta}{\alpha}}\left(-dT^2+dY^2\right)+r^2 d\Omega^2.
\label{kruskal1}
\end{eqnarray}
The metric (\ref{kruskal1}) is nonsingular at $r=r_H$. Thus ($T$,
$Y$) define a well behaved coordinate system around the event
horizon. When $r\rightarrow r_H (\approx 2M)$, from
Eq.~(\ref{kruskal3}) we have approximately (after scaling
$r\rightarrow \sqrt{\lambda} \eta r$),
\begin{eqnarray}
r\approx 2M\sqrt{\lambda}\eta+ 2M\sqrt{\lambda}\eta
\left[e^{-\frac{\beta}{\alpha}\ln\left\vert\frac{2M}{r_{C}}-1\right\vert
   - \frac{\gamma}{\alpha}\ln\left\vert
   \frac{2M}{r_U}-1\right\vert}\right]\left(Y^2-T^2\right).    
\label{kruskal4}
\end{eqnarray}
Let us expand Eq.~(\ref{eom1}) in the vicinity of the black hole
event horizon. We use Eq.~(\ref{kruskal4}) to get the following
expressions for derivatives of the scalar field $X(\rho)$
\begin{eqnarray}
\partial_{T}X&=& -4M\sqrt{\lambda}\eta
\left[e^{-\frac{\beta}{\alpha}\ln\left\vert\frac{2M}{r_{C}}-
   1\right\vert - \frac{\gamma}{\alpha}\ln\left\vert
   \frac{2M}{r_U}-1\right\vert}\right]\left(Y^2-T^2\right)  
T \partial_{\rho}X \sin \theta,\\
\partial_{Y}X&=& \hskip .3cm 4M \sqrt{\lambda}\eta
\left[e^{-\frac{\beta}{\alpha}\ln\left\vert\frac{2M}{r_{C}} -
   1\right\vert-\frac{\gamma}{\alpha}\ln\left\vert\frac{2M}{r_U} -
   1\right\vert}\right]\left(Y^2-T^2\right)
Y \partial_{\rho}X \sin \theta.
\label{derivatives}
\end{eqnarray}
Here $\rho=r\sqrt{\lambda}\eta \sin \theta$ as
before. Eq.~(\ref{eom1}) now can be written on the background
metric of Eq.~(\ref{kruskal1}) as
\begin{eqnarray}
\left(1-\sin^2\theta\right)\frac{d^2 X}{d \rho^2}
+\frac{1}{\rho}\left(1-2\sin^2\theta\right) \frac{dX}{d\rho} +
\frac{A}{8M\alpha^2 \rho \lambda\eta^2}\left\vert
\frac{r}{r_C}-1\right\vert^{\frac{\beta}{\alpha}-1}\left\vert
\frac{r}{r_U}-1\right\vert^{\frac{\gamma}{\alpha}-1}\nonumber \\
\left[2\rho^2 \frac{dX}{d\rho}+4\rho \sin\theta
\frac{dX}{d\rho}\left(r-2M \sqrt{\lambda}\eta\right)+2 \rho^2
\frac{d^2X}{d\rho^2}\sin\theta\left(r-2M
\sqrt{\lambda}\eta\right)\right]=0,
\label{horizoneq1}
\end{eqnarray}
where $A=4M\sqrt{\lambda}\eta
e^{-\frac{\beta}{\alpha}\ln\left\vert\frac{2M}{r_{C}} -
1\right\vert-\frac{\gamma}{\alpha}\ln\left\vert\frac{2M}{r_U} -
1\right\vert}$.  Under the approximations of Eq.~(\ref{approx}),
the fact that $\vert r_U \vert\approx \sqrt{\frac{3}{\Lambda}}$ and
$\left(r-2M \sqrt{\lambda}\eta\right)$ is an infinitesimal
quantity, Eq.~(\ref{horizoneq1}) reduces to Eq.~(\ref{noeq1}).
Similar arguments hold for Eq.~(\ref{eom2}).

For calculations at the cosmological horizon, we have to use the
following chart which is nonsingular as $r\rightarrow r_C $,
\begin{eqnarray}
ds^2= \frac{8M\beta^2}{r} \left\vert\frac{r}{r_U}-1
\right\vert^{1-\frac{\gamma}{\beta}}\left\vert \frac{r}{r_H}-1
\right\vert^{1-\frac{\alpha}{\beta}}\left(-{dT^{\prime}}^2 +
       {dY^{\prime}}^2\right)+r^2 d\Omega^2.
\label{kruskal5}
\end{eqnarray}
$T^{\prime}$ and $Y^{\prime}$ are timelike and spacelike
coordinates respectively, well defined around $r=r_C$. They can be
derived exactly in the same manner as for $r\approx r_{H}$.
Following a similar procedure as before one can show that Eq.s
(\ref{eom1}), (\ref{eom2}) reduce to flat space Eq.s (\ref{noeq1}),
(\ref{noeq2}) respectively. Thus the flat space equations of motion
hold on both the horizons. Since the coordinate system described in
Eq.s (\ref{kruskal1}) and (\ref{kruskal5}) are well behaved around
the respective horizons, we can also use them to expand the field
equations in regions infinitesimally beyond the horizons. For
$r\rightarrow r_H-0$ the scalar field equation (\ref{horizoneq1})
still holds. The only difference is that the quantity $\left(r-2M
\sqrt{\lambda}\eta\right)$ is negative infinitesimal. But it can be
neglected as before. Similar arguments can be given for the region
$r\rightarrow r_{C}+0$ using the chart of Eq.~(\ref{kruskal5}).

Thus with the boundary conditions on $X$ and $P$ and the
approximations of Eq.~(\ref{approx}), the configuration of matter
fields are like the Nielsen-Olesen string within, at, even slightly
beyond the horizons. Hence we conclude that a Schwarzschild-de
Sitter black hole can be pierced by a thin Nielsen-Olesen string if
the back reaction of the matter distribution to the background
space-time can be neglected.

We end with a brief remark about the backreaction of the string on
the metric. For a Schwarzschild-de Sitter space-time with a string
along the $z$-axis, the metric functions must be $z$-dependent. If
the cosmological constant were zero, the (Schwarzschild) space-time
would be asymptotically flat, and we could use Weyl
coordinates~\cite{J.L.Synge:1960zz} to write the metric in an
explicitly axisymmetric form,
\begin{eqnarray}
ds^2= -B^2 dt^2+\rho^2 B^{-2}d\phi^2+
A^2  \left[d\rho^2+dz^2\right],
\label{harmonic2}
\end{eqnarray}
where the coefficients $A, B$ are functions of $(\rho, z)$ only.
It would be relatively easy to determine the existence of cosmic
strings from the equations of motion of the gauge and Higgs fields
written in these coordinates, as was done
in~\cite{Achucarro:1995nu}. On the other hand, when the
cosmological constant is non-vanishing, it is no longer possible to
write the metric in this form. We were unable to find an
appropriate generalization of the Weyl coordinates, which are
needed to solve Einstein equations coupled to the gauge and Higgs
fields.  

We thank B.~Hartmann, S.~Deser and R.~Jackiw for reminding us of
relevant references.



\end{document}